\documentstyle[aps]{revtex}

\draft

\begin{document}
\title{Why Time is Future Oriented}
\author{Shahid N. Afridi\thanks{%
email: snafridi@phys.qau.edu.pk} and M. Khalid Khan\thanks{%
email: mkk@qau.edu.pk}}
\address{Department of Physics, Quaid-i-Azam University,\\
Islamabad, Pakistan. }
\maketitle

\begin{abstract}
We assume that the universe consists of clusters which in turns have
sub-clusters and the sub-clusters have sub-subclusters and so on. Confining
to three-dimensional space, it is shown that the universe is expanding if
entropy of the universe increases. It is also shown that clocks slow down
when time progresses towards future. We also provide a satisfactory
explanation of the big bang.
\end{abstract}


\section{Introduction}

Our physical world is the manifestation of space, time and matter. A
discipline which relates space, time and matter is the general theory of
relativity. Although general relativity is compatible with experimental
tests \cite{pdata,supernovae}, we present here an alternate approach to
incorporate matter content in a curvature-free space time.

We find an analogy between bending of light in the gravitational field and a
light beam trapped in an optical fiber. An optical fiber is able to trap a
light beam in the core by total internal reflection \cite{fiber}. Consider
an optical fiber and put it in a curved way. Let there is a light source at
one end and the observer is at the other end. The observer always sees the
source directed away from him and thus detects no curvature in the tube.
Similar is the case of a light coming from a star passing near the Sun. This
produces a slight shift in the apparent position of the star as seen by an
observer at the Earth. The observer at the Earth perceives that the star is
actually situated at its apparent position and so he connects the apparent
position of the star with the Earth according to $l=ct$.

An interesting feature of optical fibre is as follow: Take an extensionless
optical fiber. Put it in a volume element in such a way that it passes
through all points of the volume element. Send a light signal from one end
which will shine all points within the tube. The observer at the other end
sees an array of points along the tube. But since the points belong to the
volume element, thus the observer sees the whole three dimensional space in
his one dimensional tube. In other words the observer can not distinguish
between one dimension or any higher dimensions.

We incorporate this idea in an oversimplified statistical model. In this
article we emphasize on qualitative analysis, numerical estimates will be
discussed elsewhere.

\section{Model}

We assume that the universe is unique \cite{CLO}, that no part of the
universe is disconnected from the other part of the universe. We further
assume that the universe consists of super-clusters which in turns contain
group of clusters and each cluster then consists of galaxies and so on.

We can represent the universe by a set $G^{n}$

\[
G^{n}=\{G_{1}^{n{\small -}1},G_{2}^{n-1},..........,G_{p_{n}}^{n-1}\}, 
\]
where $G_{i}^{n-1}$ represents $i^{\text{th}}$ super-cluster for instant and 
$p_{n}$ is the number of super-cluster in $G^{n}$ and 
\[
G_{i}^{n-1}=\{G_{1}^{n-2},G_{2}^{n-2},..........,G_{p_{n-{\small 1}%
}}^{n-2}\}, 
\]
We can Continue this to sub-atomic level and finally we get 
\[
G_{j}^{1}=\{G_{1}^{0},G_{2}^{0},..........,G_{p_{{\small 1}}}^{0}\}, 
\]
where $G_{k}^{0}$ corresponds to constituent particle which has no further
substructure.

The number of accessible states for this system can be written as \cite
{Lee,Bowley} 
\begin{equation}
W_{n}=\left( \frac{V}{\Delta V_{n-1}}\right) ^{p_{n}},  \eqnum{1}  \label{1}
\end{equation}
where $V$ is the volume of the universe (i.e. $G^{n}$) and $\Delta V_{n-1}$
is the volume occupied by $G_{i}^{n-1}$.

Similarly, 
\begin{equation}
W_{n-1}=\left( \frac{\Delta V_{n-1}}{\Delta V_{n-2}}\right) ^{p_{n-1}}, 
\eqnum{2}  \label{2}
\end{equation}
where $\Delta V_{n-2}$ is the volume occupied by $G_{j}^{n-2}$. Putting $%
\Delta V_{n-1}$ from eq.(\ref{2}) in to eq.(\ref{1}) and on re-arranging we
get 
\begin{equation}
W_{n}W_{n-1}^{p_{n}/p_{n-1}}=\left( \frac{V}{\Delta V_{n-2}}\right) ^{p_{n}},
\eqnum{3}  \label{3}
\end{equation}

Generalizing and after re-arranging it, we finally get 
\begin{equation}
\frac{V}{\Delta V_{0}}=\prod\limits_{i=1}^{n}W_{i}^{1/p_{i}},  \eqnum{4}
\label{4}
\end{equation}
or 
\begin{equation}
\ln \left( \frac{V}{\Delta V_{0}}\right) =\sum\limits_{i=1}^{n}\frac{s_{i}}{%
k_{B}p_{i}},  \eqnum{5}  \label{5}
\end{equation}
where we have used the relation, 
\begin{equation}
s=k_{B}\ln W  \eqnum{6}  \label{6}
\end{equation}
in which $s$ is the entropy and $k_{B}$ is the Boltzmann's constant, $\Delta
V_{0}$ is the spread or uncertainty in volume of the constituent particle.
For the sake of simplicity, we assume that the volume of the universe is
spherical, i.e. $V\sim l^{3}$, and also $\Delta V_{0}\sim \Delta l^{3}$. We
get 
\begin{equation}
l=\Delta l\exp \left( S/3k_{B}\right) .  \eqnum{7}  \label{7}
\end{equation}
where 
\begin{equation}
S=\sum\limits_{i=1}^{n}s_{i}/p_{i},  \eqnum{8}  \label{8}
\end{equation}
According to the second law of thermodynomics, $S$ has to increase so as $l$%
. This means that the universe is expanding. Now according to our assumption
of hypothetical optical fiber any observer will perceive that for light ray $%
l=ct$, so eq.(\ref{7}) becomes 
\begin{equation}
t=\Delta t\exp \left( S/3k_{B}\right)  \eqnum{9}  \label{9}
\end{equation}
Writing for $S$ we have 
\[
S=3k_{B}\ln \frac{t}{\Delta t} 
\]
This determines entropy-time arrow. As from the law of increase of entropy,
entropy always increases and also time is future oriented. So we can not
draw a conclusion to say which quantity is the true independent variable,
time or entropy. In order to solve this problem we proceed as follow.

Entropy of a cluster and successive sub-cluster is related through relation 
\begin{equation}
s_{j}=p_{j}s_{j-1}  \eqnum{10}  \label{10}
\end{equation}
where $s_{j}$ is the entropy and $p_{j}$ is the number of clusters
corresponding to an element $G^{j}$. This is the difference equation which
holds provided that all elements $G^{j-1}$ of $G^{j}$ have equal entropy and
otherwise $s_{j}$ represents average entropy of \ $G^{j}$. Using eq.(\ref{10}%
) in to eq.(\ref{8}) and after iteration we finally get, 
\begin{equation}
S=N(p)s_{0}  \eqnum{11}  \label{11}
\end{equation}
where 
\begin{equation}
N(p)=1+p_{1}+p_{1}p_{2}+....+\prod\limits_{i=1}^{n-1}p_{i}  \eqnum{12}
\label{12}
\end{equation}
Therefore eq.(\ref{9}) becomes 
\begin{equation}
t=\Delta t\exp \left( N(p)s_{0}/3k_{B}\right) ,  \eqnum{13}  \label{13}
\end{equation}
in which $s_{0}$ is entropy of the constituent particle. The last equation
faces a serious problem. In order that $t$ kept on increasing then either $%
s_{0}$ should increase or $N(p)$. But since $s_{0}$ is the entropy of
constituent particle, it can not alter $t$ significantly. Therefore $N(p)$
should increase monotonically. This means that something is continuously
leaking from each element $G^{i-1}$ of $G^{i}$ in to $G^{i}$ of $G^{i+1}$.
This may correspond to some interaction which bounds the whole universe. It
also confirms that the universe is unique. Since $N(p)$ is some number which
is difficult to estimate, we can write it in terms of mass.

Like eq.(\ref{10}), we have a similar equation for mass,

\begin{equation}
m_{j}=p_{j}m_{j-1}  \eqnum{14}  \label{14}
\end{equation}
where $m_{j}$ is the total mass of $G^{j}$. Eq.(\ref{14}) holds under the
similar argument as eq.(\ref{10}). Now using eq.(\ref{14}) in to eq.(\ref{12}%
) and after iteration we finally get, 
\begin{equation}
t=\Delta t\exp \left( \lambda {\cal M}\right)  \eqnum{15}  \label{15}
\end{equation}
where 
\begin{equation}
{\cal M}=\sum\limits_{i=0}^{n-1}m_{i},  \eqnum{16}  \label{16}
\end{equation}
and 
\begin{equation}
\lambda =\frac{s_{0}}{3m_{0}k_{B}}  \eqnum{17}  \label{17}
\end{equation}
${\cal M}$ can be assumed as Lyapunov variable, a variable that never
decreases \cite{AC}. It is worth noting that spontaneous creation of matter
is discussed in literature \cite{Wheeler,Dicke} and also abandoned because
of the violation of energy-momentum conservation. It will become clear that
our case is different and quite natural.

\section{Discussion}

Using the initial condition i.e. ${\cal M}=0$ in eq.(\ref{15}) we get 
\begin{equation}
t_{0}=\Delta t,  \eqnum{18}  \label{18}
\end{equation}
where $t_{0}$ is the initial time at ${\cal M}=0$. Now putting $t_{0}$ for $%
\Delta t$ in eq.(\ref{15}) we obtain 
\begin{equation}
t=t_{0}\exp \left( \lambda {\cal M}\right)  \eqnum{19}  \label{19}
\end{equation}
This equation is interesting in the sense that zero initial time is not
allowed by the universal time scales and otherwise if we set $t_{0}=0$, then
time can never begin. Using eq.(\ref{19}) we can define time-interval, 
\begin{equation}
\Delta t=\lambda t\Delta m  \eqnum{20}  \label{20}
\end{equation}
Clearly the time interval depends on the initial time. Let us define two
intervals of time one at time $t_{i}$ and the other at time $t_{j}>t_{i}$,
given by 
\begin{equation}
\Delta t_{i}=\lambda t_{i}\Delta m_{i}  \eqnum{21}  \label{21}
\end{equation}
and 
\begin{equation}
\Delta t_{j}=\lambda t_{j}\Delta m_{j}  \eqnum{22}  \label{22}
\end{equation}
There are two alternatives:

{\rm i}) 
\begin{eqnarray*}
\Delta m_{i} &\neq &\Delta m_{j} \\
\Delta t_{i} &=&\Delta t_{j}
\end{eqnarray*}

{\rm ii}) \ 
\begin{eqnarray*}
\Delta t_{i} &\neq &\Delta t_{j} \\
\Delta m_{i} &=&\Delta m_{j}
\end{eqnarray*}
\ \ \ Let us investigate alternative ({\rm i}) first. We see that $\Delta
m_{j}<\Delta m_{i}$. Therefore ${\cal M}$ will tend to a limit and so $t$
will become constant after that $\Delta t$ will become zero. This
corresponds to close universe. On the other hand alternative ({\rm ii})
corresponds to open universe. It is clear from the second alternative that $%
\Delta t_{j}>\Delta t_{i}$. It means that clocks slow down when time
advances from present to future.

Now use 
\begin{equation}
\Delta E=\Delta mc^{2},  \eqnum{23}  \label{23}
\end{equation}
in eq.(\ref{21}) we get 
\begin{equation}
\Delta E=\frac{c^{2}\Delta t_{i}}{\lambda t_{i}},  \eqnum{24}  \label{24}
\end{equation}
which means that time advances towards future on the expense of energy.

Like time interval a similar expression can be obtained for length interval, 
\begin{equation}
\Delta l_{i}=\frac{\lambda \Delta El_{i}}{c^{2}},  \eqnum{25}  \label{25}
\end{equation}
where 
\begin{equation}
l_{i}=l_{{\small 0}}\exp \left( \lambda {\cal M}_{i}\right)  \eqnum{26}
\label{26}
\end{equation}
Eq.(\ref{25}) is also true for close universe but then the length interval
becomes zero for $l\geq l_{\max }$.

Since in our case the time interval, we call it cosmological second, is
different at different time. One is required to do care while adding
cosmological seconds from the big bang till now. The total time elapsed is,

\begin{equation}
T=\sum\limits_{k=1}^{n}\Delta t_{k}  \eqnum{27}  \label{27}
\end{equation}
where $\Delta t_{k}$ is a cosmological second at time $t_{k}$. The second
alternative enables us to discretize time,

\begin{equation}
t_{j}=t_{j-1}e^{\eta }  \eqnum{28}  \label{28}
\end{equation}
where the dimensionless increment $\eta =\lambda \Delta E/c^{2}$. \
Therefore the total time elapse is, 
\begin{equation}
T=\eta t_{{\small 0}}\sum\limits_{k=1}^{n}e^{k\eta }.  \eqnum{29}  \label{29}
\end{equation}
where $n$ represents the number of cosmological seconds. If we set $n=1$, we
get first cosmological second. For $n=2$, we get first two cosmological
seconds and so on. It is important to find the difference between two
consecutive cosmological seconds, 
\begin{eqnarray}
\Delta \tau &=&\Delta t_{n+1}-\Delta t_{n}  \eqnum{30}  \label{30} \\
&=&\eta t_{{\small 0}}e^{n\eta }(e^{\eta }-1),  \nonumber
\end{eqnarray}
which shows that the difference between two consecutive cosmological seconds
during the early stage was small. During each cosmological second an amount $%
\Delta m$ was created from energy in to mass due to vacuum fluctuation and
so an enormous amount of matter was created very rapidly at the big bang.
The creation process slowed down thereafter and it is further slowing down.

\section{Conclusion}

We have enveloped both classical and quantum regimes and treated them more
or less classically. It is a good approximation which is evident from the
expression (\ref{16}). In which the lower values of index $i$ refer to
quantum regime which contributes negligibly small as compared to the leading
terms in the expression. We have left various parameters like $\eta $ and $%
t_{0}$ undetermined. In future studies we wish to determine them

\section{Acknowledgment}

Author (SNA) would like acknowledge the ICSC- World Laboratory's financial
support.

\end{document}